\documentclass[aps,preprint]{revtex4}
\usepackage{graphicx}
\usepackage{amsmath}
\usepackage[a4paper,left=2.5cm,right=3.05cm,top=2cm,bottom=2cm]{geometry}
\begin{document}







\newcommand{\be}{\begin{equation}}
\newcommand{\ee}{\end{equation}}
\newcommand{\ben}{\begin{eqnarray}}
\newcommand{\een}{\end{eqnarray}}
\newcommand{\ra}{\rangle}
\newcommand{\la}{\langle}
\newcommand{\ov}{\overline}
\newcommand{\kn}{| n \rangle}
\newcommand{\bn}{ \langle n |}
\newcommand{\til}{\tilde}
\newcommand{\iii}{\'{\i}}

\newcommand{\pp}{p}
\newcommand{\kc}{| c \rangle}
\newcommand{\bc}{ \langle c |}
\newcommand{\kz}{| z \rangle}
\newcommand{\bz}{ \langle z|}
\newcommand{\kcz}{| \til{c}_z \rangle}
\newcommand{\bcz}{ \langle \til{c}_z |}

\newcommand{\lan}{  \lambda}
\newcommand{\kp}{| \pp \rangle}
\newcommand{\bp}{ \langle  \p2  |}
\newcommand{\ATA}{\hat{A}^\dagger \hat{A}}
\newcommand{\AAT}{  \hat{A} \hat{A}^\dagger}
\newcommand{\AT}{\hat{A}^\dagger}
\newcommand{\AO}{\hat{A}}
\newcommand{\sn}{\sum^M_{n=1}}
\newcommand{\snn}{\sum^N_{n=1}}
\newcommand{\RN}{{\cal {R}}^N}
\newcommand{\RM}{{\cal {R}}^M}
\newcommand{\kpn}{ | \psi_n \rangle}
\newcommand{\bpn}{\langle \psi_n|}
\newcommand{\kfn}{ | \phi_n \rangle}
\newcommand{\bfn}{\langle \phi_n|}
\newcommand{\kfo}{ | \ov{\phi}_n \rangle}
\newcommand{\bfo}{\langle \ov{ \phi}_n|}

\newcommand{\kf}{ | \ov{f} \rangle}
\newcommand{\kfe}{ | f^o \rangle}
\newcommand{\nn}{\nonumber \\ }
\newcommand{\nd}{\noindent}

\newcommand{\ton}{ \,;\, n=1,\ldots,N}
\newcommand{\toi}{ \,;\, i=1,\ldots,M}

\newcommand{\NC}{\hat{N}^\bot}
\newcommand{\NU}{Null(\AO)}
\newcommand{\NUC}{Null^\bot(\AO)}
\newcommand{\RA}{ Range(\AO)}
\newcommand{\PN}{\hat{P}_{N}}
\newcommand{\PNC}{\hat{P}_{N^{\bot}}}
\newcommand{\PR}{\hat{P}_{R}}

\draft
\title{Entanglement and the Lower Bounds on the Speed of Quantum Evolution}
\author{A. Borr\'as$^{1}$,  M. Casas$^{1}$,
 A.R. Plastino$^{1,\,2,\,3}$, and A. Plastino$^{2,\,3}$}

\affiliation{
$^1$Departament de F\iii sica, Universitat de les Illes Balears and IMEDEA-CSIC,
\\07122 Palma de Mallorca, Spain \\\\
$^2$Department of Physics-University of Pretoria-0002 Pretoria-South Africa \\\\
$^3$Observatory and Exact Sciences Faculty-National University La Plata-CONICET,
  C.C. 727, 1900 La Plata, Argentina
  }

\date{\today}


\begin{abstract}

The concept of quantum speed limit-time (QSL) was initially
introduced as a lower bound to the  time interval that a given
initial state $\psi_I$ may need so as to evolve into a state
orthogonal to itself. Recently [V. Giovannetti, S. Lloyd, and L.
Maccone, Phys. Rev. A {\bf 67}, 052109 (2003)] this bound has been
generalized to the case where $\psi_I$ does not necessarily evolve
into an orthogonal state, but into any other $\psi_F$. It was
pointed out that, for certain classes of states, quantum
entanglement enhances the evolution ``speed" of composite quantum
systems. In this work we provide an exhaustive and systematic QSL
study for pure and mixed states belonging to the whole
15-dimensional space of two qubits, with $\psi_F$ a not necessarily
orthogonal state to $\psi_I$. We display  convincing evidence for a
clear correlation between concurrence, on the one hand, and  the
speed of quantum evolution determined by the action of a rather
general local Hamiltonian, on the other one. \vskip 1mm
\noindent
 Pacs: 03.67.Mn, 03.67.Lx, 03.65.-w, 89.70.+c \vskip 1mm
 \noindent  Keywords: Quantum Entanglement;
 Quantum Information Theory

\end{abstract}

\maketitle \vspace{.5cm}

\newpage


\section{\bf Introduction}

One of the most fundamental concepts in the quantum description of
nature is that of entanglement
\cite{GLM03a,GLM03b,NC00,LPS98,BEZ00}, which in  recent years has
been the subject of intense research efforts (see, for instance, the
following, by no means exhaustive list of references:
\cite{GLM03a,GLM03b,NC00,LPS98,BEZ00,TLB01,BCPP02b,BCPP02a,BPCP03,BCPP05}).
A state of a composite quantum system is called ``entangled" if it
cannot be represented as a mixture of factorizable pure states.
Otherwise, the state is called separable.  Entanglement constitutes
a physical resource that lies at the heart of important  information
processes \cite{NC00,LPS98,BEZ00} such as quantum teleportation,
superdense coding, and quantum computation.

Entanglement is essential for both (i) our basic understanding of
quantum mechanics and (ii) some of its most revolutionary (possible)
technical applications. Thus, it is imperative to investigate in
detail the relationships between entanglement and {\it other}
aspects of quantum mechanics. In such a vein, particular interest is
assigned to the exploration of the role played by entanglement with
regards to  the dynamical evolution of composite quantum systems.

In this effort we will be interested in the speed up of quantum
evolution produced by entanglement. Why? Because in quantum
computation one tries to (i) avoid loss of coherence and (ii)
increase the velocity of information processing and information
transmission. Anandan and  Aharonov \cite{AA90} have shown that,
given a state $\vert  \psi \rangle$ and a curve $C$ in the
projective Hilbert space $\mathcal{P}$, the quantity

\be \label{1}  s=2\int dt \frac{\Delta E}{\hbar},   \ee with

\be \label{2} (\Delta E)^2 = \langle \psi \vert H^2 \vert  \psi
\rangle -\langle  \psi  \vert H \vert  \psi \rangle^2,\ee is
independent of the particular Hamiltonian $H$ used to transport the
state   along the curve  and is in fact the distance along $C$ as
measured by the Fubini-Study metric, deducing as a consequence  the
uncertainty relation

 \be \label{3}  (\Delta E)\Delta t \ge \frac{\hbar}{4},  \ee
where $(\Delta E)$ is the time-averaged uncertainty in energy during
the time interval $\Delta t$. {\it Equality} in Eq. (\ref{3}) holds
iff the system moves along a geodesic in $\mathcal{P}$. In this case
the evolution may be said to have minimum uncertainty, analogous to
how a Gaussian wave packet is said to have minimum position-momentum
uncertainty at a given time. More generally they define an {\it
efficiency in evolution} $\epsilon=s/s_0$, where the denominator
gives the distance along the shortest geodesic joining the initial
and final points of evolution. Loss of coherence in evolution may be
regarded as due to the time-energy uncertainty principle. In trying
to avoid such loss, speeding up evolution seem advisable.

In this regard, Margolus and Levitin \cite{ML98} have shown that the
minimum evolution time in which one state evolves to an orthogonal
one depends on the mean energy and the fluctuation. Giovannetti,
Lloyd, and Maccone \cite{GLM03a,GLM03b} recently uncover the fact
that, in certain cases, entanglement helps to ``speed up" the time
evolution of composite systems. This ``speed" of quantum evolution
is also of considerable interest because of its relevance in
connection with the physical limits imposed by the basic laws of
quantum mechanics on the {\it velocity of information processing and
information transmission} \cite{ML98,CD94,L00,KZ06}.

 The evolution ``speeding-up" ability of entanglement has been
 demonstrated {\it only} in special, if important, instances. One
 would like to ascertain that {\it it is indeed an entanglement feature},
 and not just something that happens in these instances. Thus we
 will  here undertake a general study.
 The aim of the present contribution is to make a systematic
 study of the connection between (a)
entanglement and (b) the speed of quantum evolution as determined by
the action of a rather general local Hamiltonian, by means of a
numerical simulation. Our model belongs to a family that includes
the basic models of quantum optics and cavity QED
\cite{ZG00,GLLG02,SAW03,PYUA05}. In a previous work \cite{BCPP05b} a
corresponding study was performed just for (i) pure states of
bipartite systems of low dimensionality evolving towards (ii) an
orthogonal state. Two different cases were analyzed: i) two qubits
(distinguishable) systems and  ii) bosonic or fermionic composite
(bipartite) systems of the lowest dimensionality. {\it In the
present effort} we are going to tackle an extension to the case of
two distinguishable systems of (a) pure and (b) mixed states that
evolve to (c) any other state, not necessarily orthogonal to the
initial one. We also consider (d) the special case of maximum
entangled mixed states (MEMS) \cite {MJWK01}, and also (e) that of
the set of mixed states whose entanglement degree cannot be
increased by the action of quantum gates (IH states) \cite{IH00}. We
remark on the facts that i) MEMS have recently been detected
experimentally \cite{PABJ04,BMNM04} and ii) nowadays the possibility
of obtaining such states via the action of
local non-unitary quantum channels is being studied\cite{APVW06,ZB05}. 
Thus, the ensuing results will be applicable to any physical systems
where bipartite states play a leading role.

 The paper is organized as follows: in Sec. II, using the
 time evolution of the fidelity we present the quantum speed limit
 for pure states. The case of mixed states is presented in Sec.
 III. The special case of MEMS  and IH states are presented in
 Sec. IV, and finally some conclusions are drawn in Sec. V.


\section{Quantum speed limit for pure states}

 Let us consider first  the  dynamical evolution of pure states for
 the case of  two equal but distinguishable subsystems
 evolving under a local Hamiltonian, that is, we deal with
  a two-qubits system whose evolution is governed
 by the (local) Hamiltonian

 \be \label{haloloco}
 H \, = \, H_A \otimes I_B + I_A \otimes H_B,
 \ee
 whose eigenvalue equation writes

 \ben \label{eigen0} & H_{A,B}= \epsilon_{A,B} |1\rangle, \cr
   & H_{A,B}= 0 |0\rangle, \cr
& \epsilon_A=\delta_A\epsilon, \cr &  \epsilon_B=\delta_B\epsilon, \\
& \epsilon \,\, {\rm being\,\,an\,\,arbitrary\,\,energy}, \nonumber
\een

 Our bipartite states (the eigenstates of
 $H$) are $|00\rangle$, $|01\rangle$, $|10\rangle$, and $|11\rangle$,
 while the concomitant eigenvalues equal  $0$, $\delta_B\epsilon$, $\delta_A\epsilon$,
 and $(\delta_A + \delta_B)\epsilon$, respectively.

\subsection{General methodological considerations}
In this paper we perform a systematic numerical survey of the
evolution properties of arbitrary (pure and mixed) states of a
two-qubits quantum system, under the action of the Hamiltonian
(\ref{haloloco}), by recourse to an exhaustive exploration of the
concomitant state-space ${\cal S}$. To such an end it is necessary
to introduce an appropriate measure $\mu $ on this space. Such a
measure is needed to compute volumes within ${\cal S}$, as well as
to determine what is to be understood by a uniform distribution of
states on ${\cal S}$.  The measure that we are going to adopt here
is taken from the work of Zyczkowski {\it et al.}
\cite{ZHS98,Z99}. An arbitrary (pure or mixed) state $\rho$ of a
quantum system described by an $N$-dimensional Hilbert space can
always be expressed as the product of three matrices,

\be \label{udot} \rho \, = \, U D[\{\lambda_i\}] U^{\dagger}. \ee

\noindent Here $U$ is an $N\times N$ unitary matrix and
$D[\{\lambda_i\}]$ is an $N\times N$ diagonal matrix whose diagonal
elements are $\{\lambda_1, \ldots, \lambda_N \}$, with $0 \le
\lambda_i \le 1$, and $\sum_i \lambda_i = 1$. The group of unitary
matrices $U(N)$ is endowed with a unique, uniform measure: the Haar
measure $\nu$ \cite{PZK98}. On the other hand, the $N$-simplex
$\Delta$, consisting of all the real $N$-uples $\{\lambda_1, \ldots,
\lambda_N \}$ appearing in Eq. (\ref{udot}), is a subset of a
$(N-1)$-dimensional hyperplane of ${\cal R}^N$. Consequently, the
standard normalized Lebesgue measure ${\cal L}_{N-1}$ on ${\cal
R}^{N-1}$ provides a measure for $\Delta$. The aforementioned
measures on $U(N)$ and $\Delta$ lead then to a
 measure $\mu $ on the set ${\cal S}$ of all the states of
our quantum system \cite{ZHS98,Z99,PZK98}, namely,

\be \label{memu}
 \mu = \nu {\cal L}_{N-1}.
 \ee

 \noindent

In our numerical computations we randomly generate pure and mixed
states according to the measure (\ref{memu}).

\subsection{Pure states}
 For pure states $|\Psi \rangle  $ of our composite system
 the natural measure of entanglement is the usual reduced von Neumann
 entropy $S[\rho_{A,B}] = -Tr_{A,B} (\rho_{A,B} \ln \rho_{A,B}) $
 (of either particle $A$ or particle $B$) where $\rho_{A,B} = Tr_{B,A}
 (|\Psi \rangle \langle \Psi |)$. It is convenient for our present purposes
 to use, instead of information measure $S[\rho_{A,B}]$ itself, the closely related
 ${\it concurrence \,\, value\,\,C}$, given by

 \be \label{concurre1}
 C^2=4 \det \rho_{A,B}.
 \ee

\noindent
Both the entanglement entropy $S[\rho_{A,B}]$
and the concurrence $C$ are preserved under
the time evolution determined by the local
Hamiltonian (\ref{haloloco}). Given an
initial state

\begin{equation} \label{instate}
|\Psi(t=0)\rangle = c_0 |00\rangle+c_1 |01\rangle+c_2 |10\rangle+c_3 |11\rangle,
\end{equation}
with

 \begin{equation} \label{norm}
 \sum_{i=1}^4 |c_i|^2 = 1,
 \end{equation}

\noindent its concurrence is,

\be \label{concurre2}
C^2 \, = \, 4|c_0c_3-c_1c_2|^2.
\ee

\noindent

Our objective is to characterize the departure of the system, at a
time $t$ (represented by $\Psi(t)$), from its initial state
$\Psi(t=0)$. To this end we can use the
quantum concept of fidelity $P$ that, for pure states,  is the squared-modulus of
the overlap between the two states involved, i.e.,

\begin{equation} \label{overlape}
P(z)=|\langle\Psi(t=0)|\Psi(t)\rangle|\,^2 =
|\,|c_0|^2+|c_1|^2z^{\delta_B}+|c_2|^2z^{\delta_A}+|c_3|^2z^{\delta_A+\delta_B}\,|\,^2,
\end{equation}

\noindent where \be \label{zeta} z \equiv {\rm exp}(i\Omega),\ee
 and
\be \label{Omega} \Omega=\frac{t\epsilon}{\hbar}.\ee

From now on we consider time intervals measured in  units of $\hbar
/ \epsilon$, and use for this rescaled time the letter $\Omega$. We
will also use the following notation: $\Delta^+ = \delta_A +
\delta_B$ and $\Delta^- = \delta_A - \delta_B$

The key idea is that of measuring the speed of dynamical evolution
by  studying  the time evolution of the fidelity. To such an end one
first of all fixes a given $P$ amount, say $P=F$, and proceeds to
calculate amount of time needed for  a given state  to evolve from
$P(z)=1$ (at $t=0$) to $P(z)=F$ at, say, $t=\tau$, for $F\in[0,1]$.

In Ref. \cite{BCPP05b} only the case $\delta_A=\delta_B=1~$ was
discussed. This particular Hamiltonian instance will be referred to
as corresponding to the Hamiltonian $H_I$ in Sec. IV.

\noindent The condition (\ref{overlape}) specializes for $H_I$ to

\begin{equation}\label{eqfidpure}
F \, = \, 2 p_{03} \cos{2\Omega}+ 2 p_{03} (1-s_{03}) \cos{\Omega} +
(1-s_{03})^2 + |c_0|^4  + |c_3|^4,
\end{equation}

\noindent where $p_{03} = |c_0|^2 |c_3|^2$ and $s_{03} =
|c_0|^2+|c_3|^2$.

In this case a minimum of the fidelity is achieved for the special
value $\Omega=\Omega_{min}$ given by

\begin{equation}\label{alphaminpure}
\Omega_{min}=\arccos \frac{-(1-s_{03}) s_{03} }{4 p_{03}},
\end{equation}
which can yield nonphysical complex values. To avoid this we limit
the argument of the arccos to the interval $[-1,0]$.

The time $\tau$ required to evolve to a state with fidelity $F$
admits a lower bound that depends upon both the state's expectation
energy $E$ and its fluctuation $\Delta E$ \cite{GLM03b},

\begin{equation} \label{Tmin}
T_{L.\,Bound}=max\bigg(\alpha(F)\frac{\pi\hbar}{2E},\beta(F)\frac{\pi\hbar}{2\Delta
E}\bigg),
\end{equation}

\noindent where the functions $\alpha(F)$ and $\beta(F)$ are
detailed in Ref. \cite{GLM03b}. We can compute $\beta(F)$ using an
expression previously proved in Refs. \cite{B83,P93}

\begin{equation}\label{betafid}
\beta(F)=\frac{2}{\pi}\arccos(\sqrt{F}),
\end{equation}
\noindent while $\alpha(F)$  can be numerically calculated with
great accuracy \cite{GLM03b}. For $F=0$ (when the initial state
evolves to an orthogonal one), one finds $ \alpha(F) = \beta(F) =
1$. In the opposite situation, when the state does not evolve
($F=1$), both functions vanish.

In Ref. \cite{BCPP05b} situations were dealt with for which the
orthogonal state to the initial one was definitely reached, which is
not the usual case.
 A useful parametrization,  introduced in this reference, reads

\begin{eqnarray}\label{paramorth}
|c_0|^2&=&|c_3|^2=\Gamma, \cr |c_1|^2&=&-2\delta\Gamma {\rm
cos}\Omega \cr |c_2|^2&=&-2(1-\delta)\Gamma {\rm cos}\Omega,
\end{eqnarray}

\noindent with $\Gamma=1/[2(1-{\rm cos}\Omega)]$ and $\Omega \in
[\frac{\pi}{2},\pi]$, with $\delta \in [0,1]$. In other words,
$\Omega={\rm arccos}((2\Gamma-1)/2\Gamma)$. We note that
introduction of this parametrization in the right-hand-side of Eq.
(\ref{alphaminpure}) for $F=0$ (orthogonality) does yield an
equality.


\begin{figure}
\begin{center}
\includegraphics[scale=0.35,angle=270]{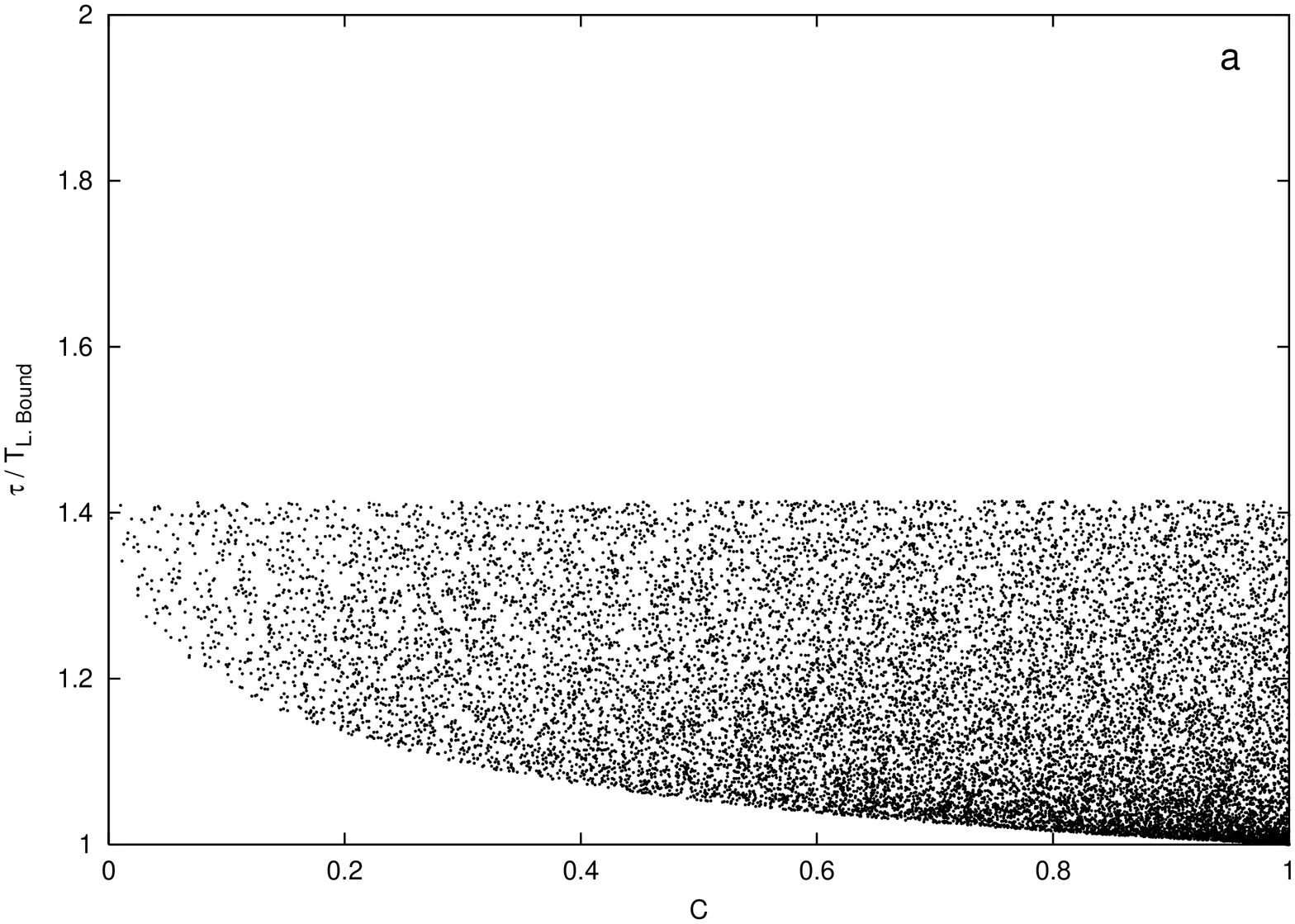}
\vspace{0.5cm}
\includegraphics*[scale=0.35,angle=270]{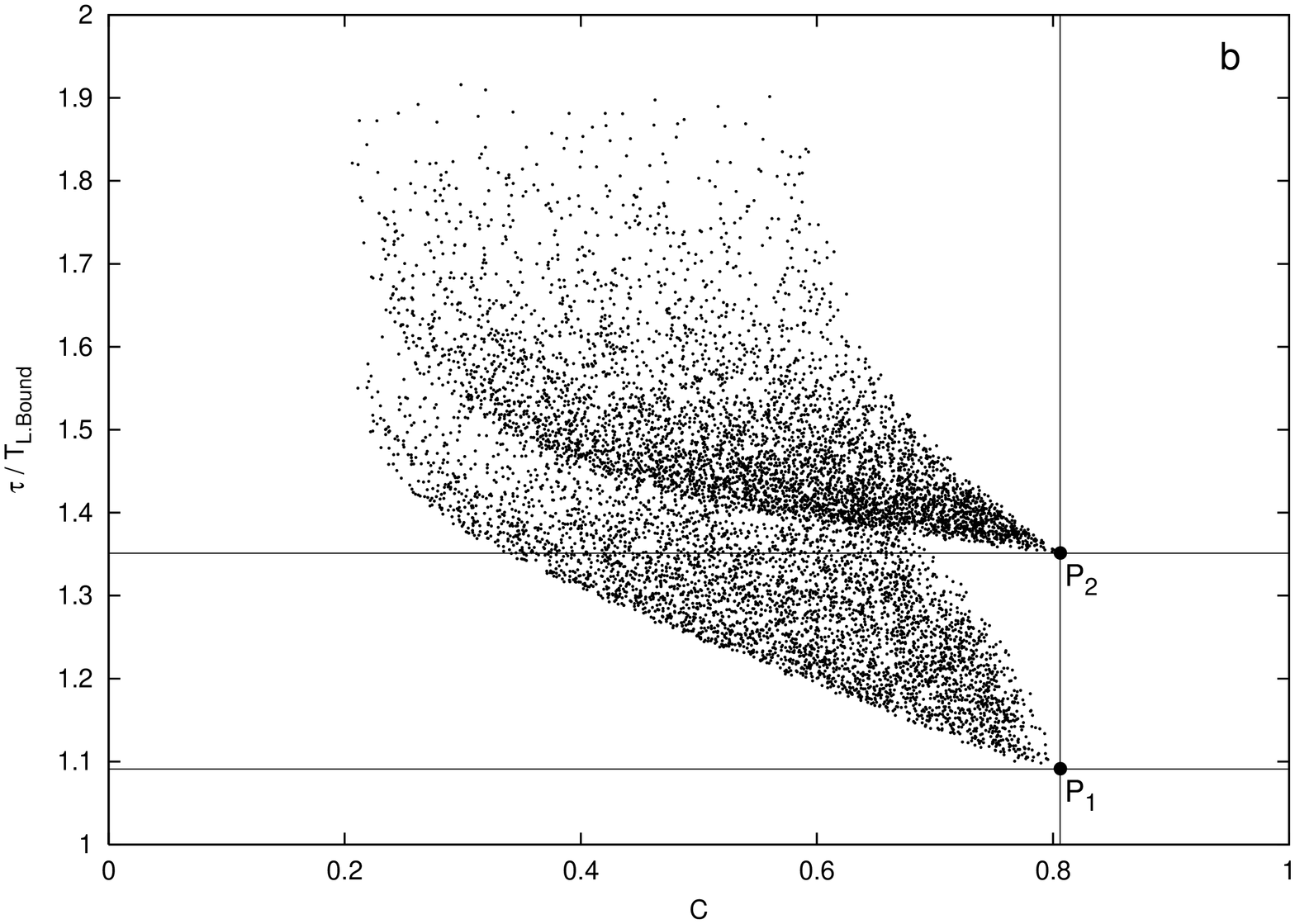}
\caption{$\tau/T_{L.\,Bound}$ for pure states. a) Pure states that
evolve to an orthogonal one and b) pure states for which $F_{min}
\in [0.35,0.4]$. The points $P_1$ and $P_2$ represent the fastest
states corresponding to each of the families determined by Eq.
(\ref{Tmin}). For the range of values of $F_{min}$ considered here,
these fast states correspond to $F_{min} = 0.35$.}
\end{center}
\label{fig1}
\end{figure}


To study the role of the entanglement on the speed of quantum
evolution and on its lower bounds, one should pay attention to the
(C,$\tau/T_{L.Bound}$) plane. A representative group of two-qubit
states evolving to an orthogonal one is depicted in Fig. 1(a). These
states can easily be generated using the parametrization
(\ref{paramorth}). The ratio $\tau/T_{L.Bound}$ has a maximum value
equal to $\sqrt{2}$, no matter which value $C$ adopts.
 The minimum value of this ratio does strongly depend
on $C$ through the value of $\Omega$, and can be analytically
obtained \cite{BCPP05b}. Only maximally entangled states reach the
bound $T_{L.Bound}$. Separable states have a different behavior. For
a rather general Hamiltonian of type $H_I$, they all evolve to an
orthogonal state in a fixed time $\tau/T_{L.Bound} = \sqrt{2}$.

These features can be easily  explained. For pure states evolving to
an orthogonal state according to $H_I$, the minimum time interval
required to complete such evolution depends {\it only} on $\Delta
E$. This is the first option in Eq. (\ref{Tmin}). In this specific
case $\Delta E$ strictly depends on the value of $\Omega$
\cite{BCPP05b}. All separable states evolve to an orthogonal one in
a fixed time $\Omega = \pi$, for which the minimum possible value of
the ratio $\tau/T_{L.Bound}$ is $\sqrt{2}$. Thus,  for separable
states the maximum and the minimum of such ratio coincide. As the
entanglement is increased, pure states are able to evolve in more
rapid fashion. This is the reason that lies behind the dependence of
the minimum of $\tau/T_{L.Bound}$ with the concurrence. Maximally
entangled states can evolve to an orthogonal state in any time lapse
within the range $\Omega \in (\frac{\pi}{2},\pi)$. The states that
evolve in the shortest possible time ($\Omega = \pi/2$) are also
those to which  a minimum of the ratio $\tau/T_{L.Bound} = 1$ is
assigned. Of course,  we cannot extend the same conclusions to the
rest of the states that evolve to some value of $F \neq 0$, as it is
clearly seen in Fig. 1(b), or to other Hamiltonians than $H_I$.

 From Eq. (\ref{eqfidpure}) we realize that
the fidelity for pure states oscillates in time. Thus, as time
goes on the fidelity of a given state climbs and goes down in
alternating fashion, reaching  minima of different depths.
 Our interest in this
respect will be focused on those special times at which {\it the
first fidelity-minimum} $F_{min}$ is attained. Doing so we can treat
all pure states in a unified manner. Note that the type of
oscillation we are speaking about has a strong dependence on the
form of the Hamiltonian one is dealing with.

We proceed to  randomly generate initial states $|\Psi \rangle  $ as explained
above. For each state we calculate: \begin{enumerate}
\item[(i)] its concurrence $C_{\Psi}$ and
\item[(ii)] the {\it first} minimum
$F_{min}^{\Psi}$ that the fidelity attains during  the time
evolution of $|\Psi \rangle  $. For a given pure state $|\Phi
\rangle$, $F_{min}^{\Phi}$ tells us ``how far" can $|\Phi \rangle$
travel in $\mathcal{S}$, before starting backwards towards itself,
as guided by the Hamiltonian. Deeper valleys may be reached later in
the periodic time evolution, but we are interested only in the one
that is reached first.
\item[(iii)] the  time $\tau$ required for the state
$|\Psi \rangle$  to reach the first fidelity minimum
$F_{min}^{\Psi}$,
\item[(iv)] the time interval $T_{L.\,Bound}$ that arises by the following process:
fixing first an {\it arbitrary} $F$ value, $|\Psi \rangle $ can
evolve to states $|\Phi \rangle$ such that the overlap
(\ref{overlape}) between $|\Psi \rangle $ and $|\Phi \rangle$
attains this value. Some time interval
$t_{o}=\Omega\,\hbar/\epsilon$ [cf. Eq. (\ref{Omega})] is needed to
reach each of these states. The minimum possible such interval is
called $T_{L.\,Bound}^{\Psi}$ and given by the bound (\ref{Tmin}).
Notice that the fidelity value $F$ reached in this time interval
needs not correspond to any  {\it minimum} of the fidelity.
\end{enumerate}

Thus, for each $|\Psi \rangle  $ we compute its concurrence
$C_{\Psi}$, the (first) minimum of the fidelity $F_{min}^{\Psi}$,
and the  time $\tau$ (in units of $T_{L.\,Bound}$), i.e., the ratio
$\tau/T_{L.\,Bound}$. This allows us to build up an association
connecting each $|\Psi \rangle  $ to these three quantities:

\be \label{mapa} |\Psi \rangle   \rightarrow \{C, F_{min},
\tau/T_{L.\,Bound} \}.\ee

A representative group of those states for which $F_{min} \in
[0.35,0.4]$ is depicted in Fig. 1(b). Their behavior is quite
different than those of Fig. 1(a). For these $F_{min}$ values there
exist two different families of states, corresponding to the two
extant possibilities for $T_{L.Bound}$ [cf. Eq. (\ref{Tmin})]. The
lower one corresponds to those states for which the bound is
determined by its expectation-energy value $E$ [the first one in Eq.
(\ref{Tmin})]. States for which the bound $T_{L.Bound}$ is
determined by its energy spread $\Delta E$ belong to the upper
group. None of these states (for both sets) reach the bound
$\tau/T_{L.Bound}=1$. The maximum value for the ratio
$\tau/T_{L.Bound}$ of Fig. 1(a) is clearly exceeded here. There also
exists a forbidden $C$ zone for states with $F_{min}$ different than
$0$. To acquire a global perspective regarding these families of
states, for any value of $F_{min}$, we will study the dependence of
their fastest evolving states [$P_1$ and $P_2$ in Fig. 1(b)] with
$F_{min}$. These {\it rapidly evolving states} achieve the minimum
of the fidelity in a time $\Omega_{min} = \pi/2$. Also, from Eq.
(\ref{eqfidpure}), we realize that the minimum of the fidelity
cannot be reached in a time shorted than $\Omega_{min} = \pi/2$. If
proper account of the normalization (\ref{norm}) of the initial
state is taken, the only compatible state parametrization turns out
to be

\begin{eqnarray}\label{coefsfidmin}
|c_1|^2 &=& |c_2|^2 = 0, \cr |c_0|^2 &=& \frac{1+\sqrt{F_{min}}}{2}
, \cr |c_3|^2 &=& \frac{1-\sqrt{F_{min}}}{2}.
\end{eqnarray}

Given such a parametrization and minding Eq. (\ref {concurre2}),
we ascertain that the concurrence  $C_{L.P_i}$ for these fast
states is the same for our two families, being completely
determined once the value of $F_{min}$ is fixed,

\begin{equation}\label{concp12}
C_{L.P_i} = \sqrt{1-F_{min}}.
\end{equation}


\begin{figure}[t]
\begin{center}
\includegraphics*[scale=0.35,angle=270]{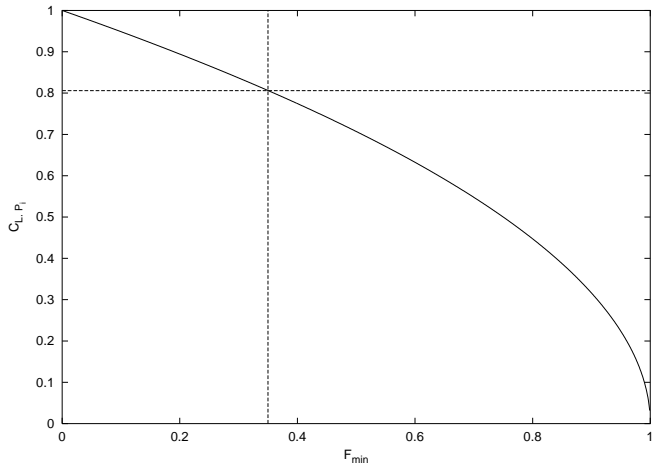}
\caption{Concurrence for the fastest pure states compatible with a
given value of $F_{min}$ as given by Eq. (\ref{concp12}). The
horizontal and vertical lines cross at the point corresponding to
the fastest states (points $P_1$ and $P_2$) of Fig. 1b. }
\end{center}
\label{fig2}
\end{figure}


This dependence on $F_{min}$ is illustrated if Fig. 2. For a given
value of $F_{min}$, the concurrence $C_{L.P_i}$ of these {\it fast
states} also coincides with the maximal concurrence value allowed
for. Thus, determining the concurrence of the fastest states
compatible with such fidelity is tantamount to finding the forbidden
$C$ zone for such $F_{min}$ value. Only in the special instance of
states capable to reach (in their evolution-trajectory) an
orthogonal counterpart (i.e., $F_{min} = 0$)  can we  obtain any
possible concurrence value. For the $F_{min}$ range depicted in Fig.
1(b), the points $P_1$ and $P_2$ correspond to $F_{min} = 0.35$.
According to Eq. (\ref{concp12}) their concurrence is $C_{L.P_i} =
0.806$, as shown in Figs. 1(b) and 2.

Once we know  the concurrence value for these special states, we
 ascertain  the time (in $T_{L.\,Bound}$ units) required to reach
$F_{min}$. We need first to compute the relation between $F_{min}$
and either the mean energy $E$ or the energy spread $\Delta E$.
Using the parametrization (\ref{coefsfidmin}) one obtains

\begin{eqnarray}\label{energyp12}
E \, &=& \, \epsilon \, ( \, 1-\sqrt{F_{min}} \, ), \cr \Delta E &=&
\, \epsilon \sqrt{1-F_{min}}.
\end{eqnarray}

Since we know that the shortest possible time needed to reach this
$F_{min}$ value is $\Omega_{min} = \pi/2$,  using (\ref{Tmin}) we
easily find the ratio $\tau/T_{L.Bound}$ corresponding to the
``fastest states" of our two families.

\begin{equation} \label{ttaup1}
\frac{\tau}{T_{L. P_1}} =
\frac{1-\sqrt{F_{min}}}{\alpha(F_{min})}
\end{equation}

\begin{equation} \label{ttaup2}
\frac{\tau}{T_{L. P_2}} =
\frac{\sqrt{1-F_{min}}}{\beta(F_{min})}
\end{equation}


\begin{figure}[t]
\begin{center}
\includegraphics*[scale=0.35,angle=270]{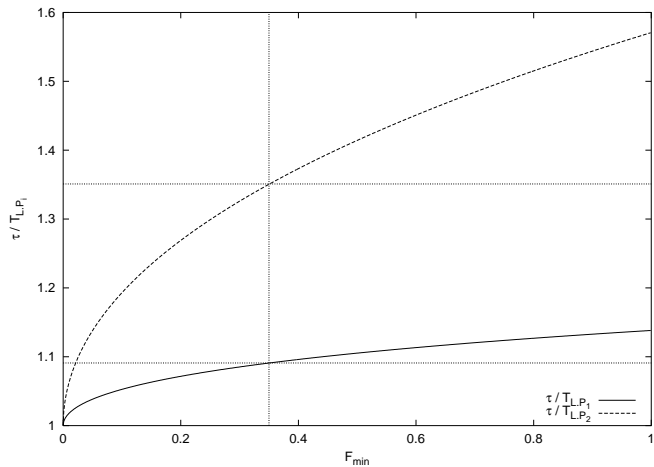}
\caption{$\tau/T_{L.\,P_i}$ for the fastest pure states compatible
with a given value of $F_{min}$ as given by Eq. (\ref{ttaup1}) for
$\tau/T_{L. P_1}$ and by Eq. (\ref{ttaup2}) for $\tau/T_{L. P_2}$.
The horizontal and vertical lines cross at two different points
corresponding to the fastest states (points $P_1$ and $P_2$) of Fig.
1(b). The upper crossing corresponds to $P_2$ and the lower one to
$P_1$.}
\end{center}
\label{fig3}
\end{figure}


Both quantities are depicted in Fig. 3. The higher the $F_{min}$
value, the more apart the two families get. They only coincide in
the $F_{min} = 0$ case, that is, for those states that evolve to an
orthogonal one. We can apply these results to the special case
considered in Fig. 1(b) ($F_{min} \in [0.35,0.4]$), remembering that
in such $F_{min}$ range the points $P_1$ and $P_2$ correspond to
$F_{min}=0.35$. For the {\it fast state} corresponding to point
$P_1$ we have $E=0.408$ (in $\epsilon$ units) and $\tau/T_{L. P_1} =
1.091$. For the state corresponding to the point $P_2$ one finds
$\Delta E=0.806$ (in $\epsilon$ units) and $\tau/T_{L. P_2} =
1.351$.

By recourse to numerical simulation we have also found that the
number of states that evolve according to Eqs. (\ref{ttaup1}) or
(\ref{ttaup2}) is a function of the value $F_{min}$. For $F_{min} =
0$ the bound for all the involved  states is given by $T_{L. Bound}
= \beta(F)(\pi\hbar/2\Delta E)$ \cite{BCPP05b}. For greater
$F_{min}$ values the situation changes.  If this value
 is large enough, approximately half of the
states belong to one of the families, while the rest are accrued to
the remaining  one.


\section{Quantum speed limit for mixed states}

Given an initial mixed state $\rho(0)$ and using the Hamiltonian of
the last section, we can easily calculate the corresponding density
matrix of the system at a given time $t$

\begin{equation}\label{genrhot}
\rho(t) = \left( \begin{array}{cccc} \rho_{11} & \rho_{12} e^{i
\delta_B \Omega} &
\rho_{13} e^{i \delta_A \Omega} & \rho_{14} e^{i \Delta^+ \Omega} \\
\rho_{21} e^{-i \delta_B \Omega} & \rho_{22} &
\rho_{23} e^{-i \Delta^- \Omega} & \rho_{24} e^{i \delta_A \Omega} \\
\rho_{31} e^{- i \delta_A \Omega} & \rho_{32} e^{i \Delta^- \Omega}
&
\rho_{33} & \rho_{34} e^{i \delta_B \Omega} \\
\rho_{41} e^{-i \Delta^+ \Omega} & \rho_{42} e^{- i \delta_A \Omega}
&
\rho_{43} e^{-i \delta_B \Omega} & \rho_{44} \\
\end{array} \right)
\end{equation}

\noindent where $\rho_{ij} = \rho_{ij}(0).$

For mixed states the fidelity's expression adopts the well known
expression

\begin{equation}\label{fidmixed}
F \big( \rho(0),\rho(t) \big) \, = \, \{ Tr [ \sqrt{\sqrt{\rho(0)}
\rho(t) \sqrt{\rho(0)}} \, ] \} ^2.
\end{equation}

In the case of pure states, this fidelity reduces to the probability
(\ref{overlape}). For the case treated here, determined by the
Hamiltonian $H_I$, the fidelity for pure states is given by Eq.
(\ref{eqfidpure}) from which we realize that it oscillates in time.
For mixed states such kind of analytical expression for the fidelity
in not available, but one can compute the fidelity numerically and
observe a similar behavior.

To study the case of mixed states we follow the same methodology
used in the previous section for pure states. Thus, we randomly
generate states $\rho$ in the two-qubits space of mixed states
$\mathcal{S}$ (of 15 dimensions). We can thus classify the values of
the ratio  $\tau/T_{L.\,Bound}$ according to their corresponding
values of the concurrence and the fidelity in such a mapping.  We
also fix our attention on the ``concurrence fidelity"-plane. Our
numerical-sampling procedure will start by constructing a fine
enough grid in the $(F , C)$ plane. We will have thus divided the
plane into a large but discrete number of ``windows". Each window,
of course, contains many states $\rho$. We will assign to all of
them the same pair of $(F , C)$ values. Notice that, for these
distinct states $\rho$, $\tau$, and $T_{L.\,Bound}$ will in general
be different. We thus average over them, but omit, for notational
simplicity's sake, the ``$<\,\,\,>$" signs. The end-result is that
we get a list of three quantities for each grid, namely,
\begin{enumerate}
\item $C$, \item $F$, \item $\tau/T_{L.\,Bound}$.
\end{enumerate}

  We  also store  information regarding the times $t$ at which the
fidelity achieves some arbitrary fixed value, not necessarily
connected with minima in any sense of the word.  Specifically, for
each $\rho$, we have selected intermediate values of the fidelity
between its initial, and maximum, value $F=1$, and its
(first-)minimum value $F_{min}^{\rho}$, according to intervals of
size $0.05$ ($F=1,0.95,0.9,0.85,\ldots$). For these fidelities, we
have stored the associated quantities $C$, $\tau/T_{L.\,Bound}$.

As stated, the time evolution for  mixed states is of a periodic
nature and the oscillation strongly depends on the Hamiltonian form.
One can circumvent to a considerable extent this $F$ dependence on
the form by using only high-fidelity values, for which, obviously,
$F$ minima cannot be reached in  arbitrarily small time intervals. A
majority of the states $\rho \in \mathcal{S}$ attain these
high-fidelity values but do not achieve, instead, lower ones. Using
high-fidelity values is then tantamount to considering most of our
randomly generated states $\in \mathcal{S}$.

Let us  focus our attention upon $\tau/T_{L.\,Bound}$. As mentioned
before, the quantity $T_{L.\,Bound}$ [cf. Eq. (\ref{Tmin})]
 (also known as the quantum speed limit time) is the lower bound for
the temporal interval required so as  to evolve, from a state $\rho$
to a state $\sigma$, in such a manner that the pair $(\rho ,
\sigma)$ of companion states corresponds to a given fidelity $F$. We
will first study this quantity for fixed fidelity values. In Fig. 4
we plot the value of $\tau/T_{L.\,Bound}$ vs the concurrence $C$ for
mixed states evolving to high fidelity companion-states. There
exists a clear correlation between the quantum speed evolution time
and the concurrence. The more entangled  a state is, the less time
it takes to reach a companion state such that the generalized
overlap between them is $F$. Indeed, this time is seen to be close
to the limit $T_{L.\,Bound}$ for high $C$ values. The relation
between $\tau/T_{L.\,Bound}$ and the concurrence $C$ does not seem
to strongly  depend on the specific $F_{min}$ value considered.
Contrariwise, for pure states it is only when we consider  {\it
small} fidelity values that a clear correlation between
$\tau/T_{L.\,Bound}$ and the concurrence $C$ is observed. These
small fixed values of the fidelity are obviously very close to its
corresponding $F_{min}$ value. Remind that we have shown in the
previous section that the correlation between $\tau/T_{L.\,Bound}$
and the concurrence $C$ does exist for $F_{min}$. If one selects
pure states with high fixed values of the fidelity $F$, the
correlation we are here speaking about is not detected, as
$\tau/T_{L.\,Bound}$ is approximately constant and close to unity
for all possible values of the concurrence $C$.


\begin{figure}[t]
\begin{center}
\includegraphics*[scale=0.35,angle=270]{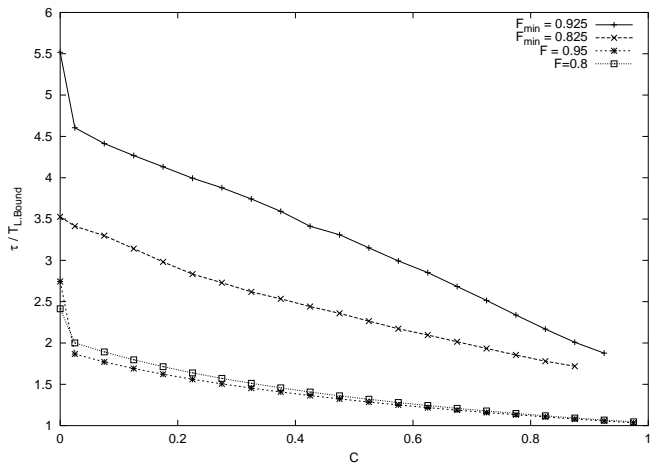}
\caption{$\tau/T_{L.\,Bound}$ for mixed states that evolve to
several fixed ($F = 0.95$ and $F= 0.8$) and minimum ($F_{min} =
0.925$ and $F_{min}= 0.825$) values of the fidelity. See text for
details. }
\end{center}
\label{fig4}
\end{figure}


Thus far we have considered arbitrary fidelities, not necessarily
linked to minima of this quantity. In Fig. 4 we also consider time
intervals $\tau$ needed to reach the first fidelity minimum. The
resulting situation resembles the one just described above. The
number of highly entangled ($C>0.9$) mixed states evolving to
companion states with a high value of the fidelity ($F>0.75$) is
very small. This entails  that we cannot numerically obtain enough
states in this zone to perform our averaging procedure in a reliable
manner. Although the limit value $\tau/T_{L.\,Bound} = 1 $ is not
reached for maximally entangled states, mixed states exhibit also in
this case a nitid speed-concurrence correlation. Consequently, their
evolution speed strongly depends on entanglement degree.


\section{Maximally Entangled Mixed States (MEMS) and IH States}

As we are interested in putative relations between entanglement and
the speeding up of the quantum evolution, the study of some special
types of  states should be of interest. In particular, the so-called
MEMS \cite{MJWK01} are  states that have the maximum possible amount
of entanglement of formation. We remark on the fact that MEM states
have recently been encountered in the laboratory
\cite{PABJ04,BMNM04,APVW06}. The associated density matrix is
written in terms of a variable $x$ that ranges in $[0,1]$. In the
basis referred to in the Sec. II their representative matrices read

\begin{equation}\label{rhomems}
\rho_{MEMS} = \left( \begin{array}{cccc} g(x) & 0 & 0 & x/2 \\
0 & 1-2g(x) & 0 & 0 \\ 0 & 0 & 0 & 0 \\ x/2 & 0 & 0 & g(x) \\
\end{array} \right),
\end{equation}
\noindent with $g(x)=1/3$ for $0 \le x \le 2/3$, and $g(x)=x/2$ for
$2/3 \le x \le 1$.

Also of great interest are the so-called Ishizaka and Hiroshima (IH)
states \cite{IH00}, whose entanglement degree cannot be increased by
acting on them with logic gates. Of course, MEMS are a special
instance of the IH class. The associated $\rho_{IH}$ matrices, of
eigenvalues $p_i;\,\,\,(i=1,2,3,4)$, read

\begin{equation}\label{rhoih}
\rho_{IH} = \left( \begin{array}{cccc} p_2 & 0 & 0 & 0 \\
0 & \frac{p_3+p_1}{2} & \frac{p_3-p_1}{2} & 0
\\ 0 & \frac{p_3-p_1}{2} & \frac{p_3+p_1}{2} & 0 \\ 0 & 0 & 0 & p_4 \\
\end{array} \right),
\end{equation}
\noindent where the eigenvalues are size ordered: $ p_1 \ge p_2 \ge
p_3 \ge p_4$. If one compares Eqs. (\ref{rhoih}) and (\ref{genrhot})
it is easy to see that IH states can ``evolve" only  if $\delta_A
\neq \delta_B$. This entails that we cannot use here  the same
Hamiltonian ($H_I$) employed in the preceding sections (see Sec.
II). For IH states we will use the values $\delta_B=1$ and
$\delta_A=2$, namely, we employ a local Hamiltonian $H_{II}$, whose
diagonal is {0,$\epsilon$,$2\epsilon$,$3\epsilon$}. For MEM states
we will also use this Hamiltonian because we want to compare its
associated results with those of the IH states. If one uses the
hamiltonian $H_I$ with MEM states the ensuing results resemble those
of the preceding section. Thus, comparison can be made with the
results of mixed states detailed there. In the MEMS instance the
fidelity can readily be computed and reads

\begin{eqnarray}\label{fidmems}
F_{MEMS} \, &=& \, \Bigg(A + \frac{1}{2}
(\sqrt{B+\sqrt{C}}+\sqrt{B-\sqrt{C}}) \Bigg)^2,
\end{eqnarray}

\noindent with

\begin{eqnarray}\label{parsfidmems}
A \, &=& \, 1-2g(x), \cr B \, &=& \, 4 g(x)^2 + x^2 \cos{(\Delta^+
\alpha)}, \cr C \, &=& \, x^2(\cos{(\Delta^+ \alpha)}-1) \Big(
8g(x)^2+ x^2(\cos{(\Delta^+ \alpha)}-1) \Big).
\end{eqnarray}

The MEMS's fidelity  expression (\ref{parsfidmems}) also applies,
with different coefficients, to the IH states case. Their
corresponding coefficients are

{\setlength\arraycolsep{1pt}
\begin{eqnarray}\label{parsfidih}
A_{IH} &=& p_2+p_4, \cr B_{IH} &=& (p_1+p_3)^2 + (p_1-p_3)^2
\cos{(\Delta^- \alpha)}, \cr C_{IH}  &=&  \Big( 1+\cos{(\Delta^-
\alpha)} \Big) \bigg( ({p_1}^2 + {p_3}^2 ) \Big( 1+\cos{(\Delta^-
\alpha)} \Big) \nonumber
\\&&{} + 2p_1p_3 \Big( 3-\cos{(\Delta^- \alpha)} \Big)
\bigg).
\end{eqnarray}}

The oscillating part of the MEMS fidelity seems to depend on
$\cos{(\delta_A+\delta_B)\Omega}$ and the minimum fidelity value
 coincides with the minimum of its oscillating part, i.e.,
the minimum fidelity is achieved at $\Omega^{MEMS}_{min} =
\pi/(\delta_A+\delta_B)$. For the IH-states the situation is
similar, the fidelity depends on $\cos{(\delta_A-\delta_B)\Omega}$
and its first minimum is achieved at $\Omega^{IH}_{min} =
\pi/(\delta_A-\delta_B)$.

We can also obtain an analytic expression for the expectation value
of the hamiltonian $E$ and its fluctuation $\Delta E$ in the case of
the MEMS states. We find

{\setlength\arraycolsep{0pt}
\begin{eqnarray} \label{energymems}
E_{MEMS} \, &=& \, \epsilon \, \big( \delta_B + g(x) \Delta^{-} \big), \cr \Delta E_{MEMS} \, &=& \, \epsilon \,
\sqrt{g(x)({\delta_B}^2+{\delta_A}^2)- {g(x)}^2
{(\Delta^{-})}^2}.
\end{eqnarray}}

For the Hamiltonian considered here  $E$ is always greater than
$\Delta E$. According to Eq. (\ref{Tmin}) (see also the paragraph
following it) one looks for the maximum of a pair of quantities.
Here $T_{L.\,Bound}$ is always equal to that one depending on
$\Delta E$, because $\beta(F)$ is always greater than or equal to
$\alpha(F)$. Taking into account
 all these results, the  equation for $\tau$ turns out to read

\begin{equation}\label{ttaumems}
\frac{\tau}{T_{L.\,Bound}}=\frac{\Omega_{min}\Delta E}{\epsilon
\arccos{\sqrt{F}}},
\end{equation}
and one must substitute $\Delta E$ and $F$ with their  pertinent
associated values, depending on (i) which zone we are working in and
(ii) which Hamiltonian we are referring to. For $\Omega_{min}$ we
obtain the values $\pi/2$, for ($\delta_B=1$ , $\delta_A=1$), and
$\pi/3$, for ($\delta_B=2$ , $\delta_A=1$).

For the IH states we obtain the following equations for $E$ and
$\Delta E$

\begin{eqnarray} \label{energih}
E_{IH} \, &=& \, \frac{\epsilon}{2} \Big( p_1 \Delta^- + (p_3+2p_4)
\Delta^+ \Big), \cr \Delta E_{IH} \, &=& \, \frac{\epsilon}{2} \sqrt{
2 p_1 \Delta^+ \Delta^- + 2(p_3+2p_4) {(\Delta^+)}^2 - \Big( p_1
\Delta^- + (p_3+2p_4) \Delta^+ \Big)^2}.
\end{eqnarray}

Notice that the MEM states are completely determined by the
parameter $x$, which corresponds to the value of the concurrence ($C
= x$). For a given value of the concurrence $C$ there exists only
one value for that magnitudes we are interested on: {$F_{min}$, $E$,
$\Delta E$ and $T_{L. Bound}$}. This means that if we want to
analyze these states we must do it using $F_{min}$ instead of fixed,
arbitrary values of the fidelity. If we use {\it fixed, arbitrary
fidelity values} $K$, one can always detect  a range of $C$ values
for which no MEMS characterized by $K$ exist. This feature
constitutes a great difference with respect to what happens for the
general mixed states case discussed above, where one has many
different states compatible with a given value of the concurrence
$C$. Such  is the case for IH states, for which we can average such
magnitudes as we did earlier for general mixed states in Sec. III.
We must mind this  difference between MEMS and IH states if we want
to compare MEM results to IH ones. As stated before, for an
arbitrary MEM state the concurrence, say $C=K$ is fixed, and so is
the value of $F_{min}$. But for the same $K$ value there exist many
IH states characterized by a wide range of possible $F_{min}$
values.


\begin{figure}[t]
\begin{center}
\includegraphics*[scale=0.35,angle=270]{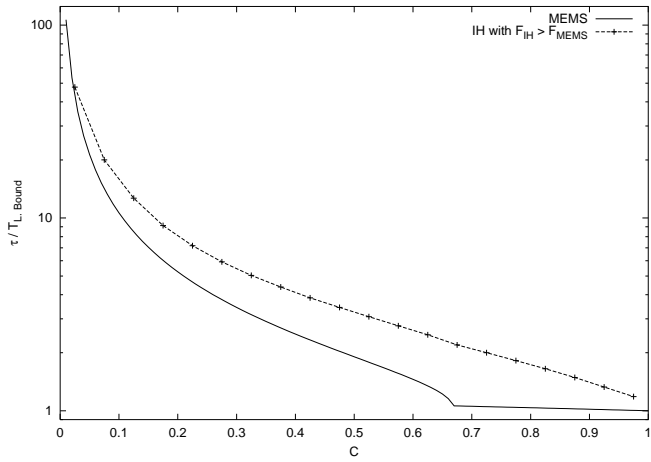}
\caption{$\tau/T_{L.\,Bound}$ for MEMS with hamiltonian $H_{II}$ and
the corresponding average for the IH case. See text for details}
\end{center}
\label{fig5}
\end{figure}


In order to be able to compare IH and MEM states we have used the
following criterion:  consider those IH states of concurrence $C=K$
with an $F_{min}$ value greater than that pertaining to the
associated $K$ MEM state. In Fig. 5 we depict $\tau/T_{L.\,Bound}$
vs $C$ for MEMS and for those IH states that fulfill the above
criterion. The behavior of separable states, with exceedingly large
$\tau/T_{L.\,Bound}$ values, is very different from that of highly
entangled ones. The ensuing differences are larger  than for the
general mixed states studied in the preceding section. There is a
clear difference between the two MEM zones arising out of the
$x-$MEM parametrization. For weakly entangled states,
$\tau/T_{L.\,Bound}$ achieves very high values. For highly entangled
states (the second MEM zone), the situation is the opposite. In this
last zone $\tau/T_{L.\,Bound}$ tends to saturate its lower bound.
The IH states have a similar behavior than the MEMS, although its
corresponding average of the ratio $\tau/T_{L.\,Bound}$ is always
greater than the corresponding MEMS ratio.

\section{Conclusions}

For mixed states of bipartite systems ruled by a general local
Hamiltonian we have put forward rather convincing evidence of the
clear correlation extant between concurrence and speed of time
evolution. The more entangled an initial state $\rho_1$ is, the less
time (in units of $T_{L. Bound}$) it takes to evolve to another
state $\rho_2$, no matter what the $\rho_1$ -$ \rho_2$ fidelity is.
In the case of pure states the correlation is strong for states that
evolve to a minimum of the fidelity. For some special mixed states,
namely, the so-called maximally entangled mixed states, the
correlation between concurrence and the speed of time evolution
becomes more acute than in the case of general mixed states.

\vskip 3mm {\bf Acknowledgements} \vskip 3mm This work was
partially supported by the MEC grant FIS2005-02796 (Spain) and
FEDER (EU) and by CONICET (Argentine Agency). A. Borr\'as
acknowledges support from the FPU grant AP-2004-2962 (MEC-Spain).

\end{document}